\documentclass[prb,twocolumn,superscriptaddress,showpacs]{revtex4-1}

\usepackage{graphicx}
\usepackage{dcolumn}
\usepackage{bm}
\usepackage{subfigure}
\usepackage{amsmath}

\begin{document}

\preprint{RAE-LaSrCoO3-1}

\title{Ferromagnetic Excitations in La$_{0.82}$Sr$_{0.18}$CoO$_{3}$ Observed Using Neutron Inelastic Scattering}

\author{R. A. Ewings}
\email{russell.ewings@stfc.ac.uk} \affiliation{ISIS Facility,
Rutherford Appleton Laboratory, STFC, Chilton, Didcot, Oxon, OX11
0QX, United Kingdom} \affiliation{ Department of Physics,
University of Oxford, Clarendon Laboratory, Parks Road, Oxford, OX1
3PU, United Kingdom}
\author{P. G. Freeman} \affiliation{Institut Laue-Langevin, BP 156, 38042 Grenoble CEDEX 9, France}
\author{M. Enderle} \affiliation{Institut Laue-Langevin, BP 156, 38042 Grenoble CEDEX 9, France}
\author{J. Kulda} \affiliation{Institut Laue-Langevin, BP 156, 38042 Grenoble CEDEX 9, France}
\author{D. Prabhakaran} \affiliation{ Department of Physics,
University of Oxford, Clarendon Laboratory, Parks Road, Oxford, OX1
3PU, United Kingdom}
\author{A. T. Boothroyd} \affiliation{ Department of Physics,
University of Oxford, Clarendon Laboratory, Parks Road, Oxford, OX1
3PU, United Kingdom}

\date{\today}

\begin{abstract}
Polarized neutron inelastic scattering has been used to measure spin excitations in ferromagnetic La$_{0.82}$Sr$_{0.18}$CoO$_{3}$. The magnon spectrum of these spin excitations is well defined at low energies but becomes heavily damped at higher energies, and can be modeled using a quadratic dispersion. We determined a spin wave stiffness constant of $D=94\pm 3$\,meV\,\AA$^{2}$. Assuming a nearest-neighbor Heisenberg model we find reasonable agreement between the exchange determined from $D$ and the bulk Curie temperature. Several possible mechanisms to account for the observed spin-wave damping are discussed.

\end{abstract}

\pacs{75.47.Lx, 75.47.Gk, 75.30.Ds}

\maketitle

\section{Introduction}\label{s:LSCoO intro}

The most studied colossal-magnetoresistive (CMR) perovskites to
date have the general formula La$_{1-x}$A$_{x}$MnO$_{3}$, where A
is a hole dopant such as Sr$^{2+}$, Ca$^{2+}$, Ba$^{2+}$, etc \cite{CMR review,CMR theory book}. These materials have a property known as phase
separation, which means spatially separated regions exist with different magnetic properties. Phase separation is believed to be important for the CMR effect.

Materials with the general formula La$_{1-x}$Sr$_{x}$CoO$_{3}$ (LSCoO) also display many of the characteristics typical of the CMR perovskites, in particular magnetic phase separation \cite{Wu CMR paper,He bulk meas prb}. Bulk measurements of the magnetic and transport properties of LSCoO show dramatic changes with doping \cite{Kriener CaSrBa,Wu glass phase}. The undoped material is a non-magnetic semiconductor, but as doping is increased the bulk magnetization increases rapidly, concomitant with a steady decrease in the resistivity, and a transition into a true ferromagnetic state between $x=0.18$ and $x=0.22$. The Curie temperature ($T_{\rm C}$) at $x=0.18$ is 150\,K, and $T_{\rm C}$ increases with further increase in doping. In the ferromagnetic phase the magnetic easy axis has been found to be the $[100]$ direction in the rhombohedral unit cell \cite{Caciuffo paper}, which corresponds to the $[110]$ direction in pseudo-cubic notation (we use pseudo-cubic notation throughout this paper, see appendix for details). For $0 < x \leq 0.18$ the material displays behavior which is in some respects like that of a spin glass \cite{Wu glass phase}, arising from the growth, and then percolation at $x=0.18$, of ferromagnetic clusters. The growth of the magnetization with doping is thus due to an increase in the number of magnetic sites contained within the clusters. It has been suggested that inter-cluster interactions give rise to the spin-glass-like behavior, whereas the intra-cluster interactions give rise to the decreasing resistivity and increasing magnetic moment. NMR \cite{Kuhns NMR} and SANS measurements \cite{Wu SANS} can also be explained using the cluster model. For $0.18\leq x \leq 0.22$ there is still some evidence of phase separation between ferromagnetic and non-magnetic clusters \cite{He bulk meas prb}, i.e. long-range ferromagnetic order exists but the ferromagnetic phase fraction is $<100\%$.

The magnetoresistance (MR) of LSCoO as a function of doping has also been measured \cite{Wu CMR paper,Aarbogh_CMR transport}. For higher dopings, when the material is in a ferromagnetic metallic phase, the MR is just a
few percent. However when the doping reaches the critical level of
$x=0.18$ the MR is $\sim 30$\%, and as the doping is
decreased further the MR increases such that for $x=0.09$ the
resistivity drops by $\sim 90$\% upon application of a magnetic field of 90\,kOe at low temperatures.

Studies of lightly doped ($x\approx0.002$) LSCoO with muons \cite{Sean muons} and with neutron inelastic scattering \cite{Podlesnyak light doping}
have shown that some of the underlying `matrix' of Co$^{3+}$ ions (i.e. LaCoO$_{3}$) changes from a non-magnetic ($S=0$, t$_{2g}^{6}$) to a magnetic
state ($S=1$, t$_{2g}^{5}$\,e$_{g}^{1}$) upon hole doping, since the size of the magnetic moment is too large to be accounted for by the small number
of Co$^{4+}$ spins ($S=1/2$, t$_{2g}^{5}$) present at such low doping levels. It is reasonable to assume that this scenario persists at higher doping,
in which case double exchange would tend to favor ferromagnetic correlations between Co$^{3+}$ and Co$^{4+}$ ions. Neutron inelastic scattering
measurements of the $x=0.1$ material \cite{Phelan PRL droplets} tend to support this double exchange interpretation, and the large reduction in resistivity as doping is increased can also be understood in these terms, together with transport between ferromagnetic clusters \cite{Aarbogh_CMR transport}.

There have been no previous microscopic studies of the ferromagnetic phase ($x\geq0.18$) of LSCoO, so although measurements of LSCoO with lighter doping indicate that double exchange is the mechanism by which LSCoO becomes ferromagnetic, there have been no direct measurements of the ferromagnetic phase to confirm this. This is in contrast to the manganites, where the double exchange mechanism has been confirmed to apply over a wide range of doping fractions, and which have been extensively studied using neutron scattering \cite{Ye many manganites,Fernandez-Baca LCMO,Moussa everyone wrong}. Determining whether this is also the case for LSCoO will be an important step towards determining whether the MR effects observed in hole-doped LSCoO have the same origin as those in the hole-doped manganites.

Despite the similarities between the cobaltites and manganites, there are good reasons to expect their neutron inelastic scattering spectra to be different. For example, recent neutron scattering studies have uncovered static incommensurate magnetic order \cite{Phelan ICM,Phelan PRL satellites} in cobaltites with $x<0.18$, and that this competes with the ferromagnetism with which we concern ourselves in this study. Furthermore, in the manganites the Mn$^{3+}$ and Mn$^{4+}$ ions have t$_{2g}^{3}$\,e$_{g}^{1}$ and $t_{2g}^{3}$
configurations respectively, whereas in the cobaltites the Co$^{3+}$ ions are in the t$_{2g}^{6}$\,e$_{g}^{0}$, t$_{2g}^{5}$\,e$_{g}^{1}$, or
t$_{2g}^{4}$\,e$_{g}^{2}$ states, and the Co$^{4+}$ is in the
t$_{2g}^{5}$ configuration. The manganites are therefore orbitally
ordered, whereas the degeneracy of the spin states in the
cobaltites is greater, a difference which may well contribute to differences in the MR properties, and also the neutron scattering spectra.

Here we report neutron inelastic scattering measurements on ferromagnetic La$_{0.82}$Sr$_{0.18}$CoO$_{3}$ in an applied magnetic field.  We characterized the lower energy excitations ($E \lesssim 20$\,meV) in terms of a quadratic dispersion, corresponding to a spin-wave stiffness of $D=94\pm 3$\,meV\,\AA$^{2}$. At higher energies, at
wavevectors closer to the Brillouin zone boundary, the scattering from the spin excitations became too weak to measure, probably due to increased
damping. Comparison of the lower energy parameters appear to be in broad agreement with a Heisenberg model in which there are nearest neighbor
interactions only.

\section{Experimental details}\label{s:LSCoO expt details}

\begin{figure}[!h]
\includegraphics*[scale=0.38,angle=270]{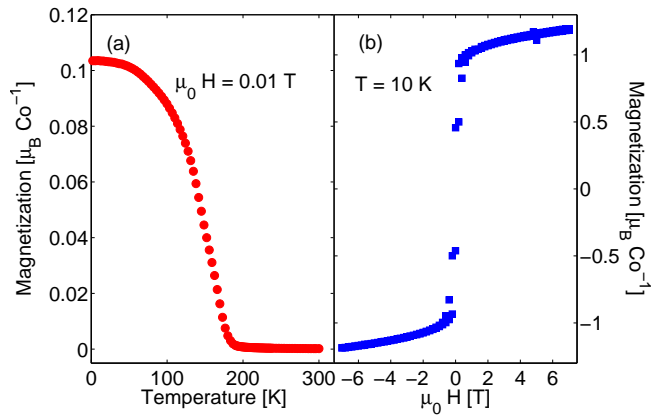}
\centering \caption{Magnetization of La$_{0.82}$Sr$_{0.18}$CoO$_{3}$ (a) as a function of temperature while cooling in a magnetic field of 0.01\,T
applied parallel to the [001] direction; (b) as a function of applied field at $T=10$\,K.}\label{fig:LaSrCoO3 squid}
\end{figure}

A single-crystal of La$_{0.82}$Sr$_{0.18}$CoO$_{3}$ of mass 8.8\,g
was grown by the floating-zone method in an image
furnace\cite{Prabhak crystal}. A small off-cut of this single-crystal was checked in a Quantum Design MPMS SQUID magnetometer, and its magnetization
as a function of temperature and applied magnetic field, shown in figure \ref{fig:LaSrCoO3 squid}, were found to be in agreement with previous
measurements \cite{Kriener CaSrBa}.

For the neutron scattering experiments, the larger part of the crystal, which was a cylinder of length 40\,mm and diameter 8\,mm, was mounted in a
horizontal-field cryomagnet and aligned with the pseudo-cubic $[100]$ and $[011]$ directions in the horizontal scattering plane. The crystal was cooled from room
temperature to 2\,K in a magnetic field of 3.5\,T. The field was
applied parallel to the pseudo-cubic $[011]$  magnetic easy direction to align the ferromagnetic domains.

All of the neutron scattering experiments were carried out at the Institut
Laue-Langevin on the IN20 triple-axis spectrometer configured for
polarization analysis. The incident neutron energy was
selected by Bragg reflection from either a Heusler or a silicon
monochromator depending respectively on whether polarized or unpolarized neutrons were required. A Heusler analyzer was used to select the energy and the polarization state of the scattered neutrons. Scans were performed with a fixed final energy $E_f$ of either 14.7 or 34.8 meV. A pyrolytic graphite (PG)
filter was present to suppress higher-order harmonics in the
scattered beam. Mezei-type spin flippers \cite{Mezei flipper} were placed in both the incident and scattered beams when the Heusler monochromator was
used, and just in the scattered beam when the silicon monochromator was used.

In order to reduce the large stray fields of the horizontal magnet at the flipper positions without increasing the monochromator-sample and sample-analyzer distances, we performed the measurements at a field of 1\,T after having cooled the sample to 2\,K in an applied field of 3.5\,T. The reduction in magnetization, and hence domain alignment, is only about 10 \% (see fig. \ref{fig:LaSrCoO3 squid}) when going from 3.5\,T to 1\,T.

The remaining stray fields at the spin flipper positions were compensated by additional coils. The compensation currents were adjusted as a function of the orientation of the horizontal magnet with respect to the incident or scattered beam. On the direct beam, flipping ratios of 8 for $E_{i}=E_{f}=34.8$\,meV and 17 for $E_{i}=E_{f}=14.7$\,meV were typical.

Scans were performed either as a function of energy $E$ at fixed
scattering vector $\bf Q$, or as a function of $\bf Q$ at fixed $E$.
During the scans, $\bf Q$ was constrained to the line $(0,1+q,1+q)$
in reciprocal space so that at all times $\bf Q$, the magnetic field $\bf H$, and hence the sample magnetization $\bf M$ and the neutron polarization $\bf P$ were parallel to one another.

In a magnetic field, the neutron polarization is resolved along the
magnetic field direction. In this uniaxial geometry, the scattering can be represented by four partial cross sections corresponding to the two possible neutron spin states before and after scattering. We will denote these by
$\uparrow\uparrow$, $\downarrow\downarrow$, $\uparrow\downarrow$ and
$\downarrow\uparrow$. The polarization state of the incident and of the scattered neutrons is controlled by the monochromator and incident beam spin
flipper, and the analyzer and scattered beam spin flipper respectively. The application of uniaxial polarization analysis to neutron scattering from a
ferromagnet is described in detail by Moon {\it et al.}\cite{Moon-PR-1969}. For the ${\bf P} \parallel {\bf Q} \parallel {\bf M}$ geometry used here,
the partial cross sections are proportional to the response functions

\begin{eqnarray}
S_{\uparrow\uparrow}({\bf Q},E) & = & S_{\rm coh}^{\rm Nuc} + S_{\rm
inc}^{\rm iso} + \mbox{$\frac{1}{3}$}S_{\rm inc}^{\rm sp}\nonumber\\[5pt]
S_{\downarrow\downarrow}({\bf Q},E) & = & S_{\rm coh}^{\rm Nuc} +
S_{\rm inc}^{\rm iso} + \mbox{$\frac{1}{3}$}S_{\rm inc}^{\rm sp}\nonumber\\[5pt]
S_{\uparrow\downarrow}({\bf Q},E) & = & \mbox{$\frac{2}{3}$}S_{\rm inc}^{\rm sp} + S_{-+}^{\rm M}\nonumber\\[5pt]
S_{\downarrow\uparrow}({\bf Q},E) & = & \mbox{$\frac{2}{3}$}S_{\rm
inc}^{\rm sp} + S_{+-}^{\rm M},\label{eq:Sxx_partial_XS}
\end{eqnarray}

where $S_{\rm coh}^{\rm Nuc}$ is the nuclear coherent scattering,
$S_{\rm inc}^{\rm iso}$ and $S_{\rm inc}^{\rm sp}$ are the nuclear
isotopic and spin incoherent scattering, and $S_{-+}^{\rm M}$ and
$S_{+-}^{\rm M}$ are response functions describing the transverse magnetic correlations. For a ferromagnet, $S_{-+}^{\rm M}$ and $S_{+-}^{\rm M}$ describe scattering processes in which magnons are created and annihilated, respectively. At temperatures such that $k_{\rm B}T \ll E$, as applicable here, only magnon creation processes ($S_{-+}^{\rm M}$) have significant
intensity.


With a half-polarized setup, in which the incident neutron beam is unpolarized but the polarization state of the neutrons in the scattered beam is analyzed, there are only two partial cross sections, corresponding to the two final polarization states. These are related to the four partial cross sections in
eq.~(\ref{eq:Sxx_partial_XS}) by $S_{\circ\uparrow} =
(S_{\uparrow\uparrow}+S_{\downarrow\uparrow})$ and
$S_{\circ\downarrow} = (S_{\downarrow\downarrow}+S_{\uparrow\downarrow})$. Although $S_{\circ\uparrow}$ and $S_{\circ\downarrow}$ contain both nuclear and magnetic scattering, the magnetic scattering can be isolated by subtraction. Specifically, from (\ref{eq:Sxx_partial_XS}) the magnon creation scattering is given by
\begin{equation}
S_{-+}^{\rm M} = S_{\circ\downarrow} - S_{\circ\uparrow}.\label{eq:half-pold}
\end{equation}
The response functions are related to the imaginary part of the
generalized susceptibility $(\chi_{\alpha \beta}({\bf
Q},E))$ through the fluctuation--dissipation
theorem
\begin{equation}
S_{\alpha\beta}({\bf Q},E) =
\{1+n(E)\}\frac{1}{\pi}\chi_{\alpha\beta}''({\bf
Q},E),\label{eq:fluctuation_dissipation_thm}
\end{equation}
where $n(E)=[\exp(E/k_{\rm B}T)-1]^{-1}$.

A few scans were performed with the fully polarized setup (i.e. Heusler monochromator and analyzer), which allows one to measure the four cross-sections detailed in eq. \ref{eq:Sxx_partial_XS}. However, as already stated, the majority of our measurements were made with a half-polarized setup in which the final neutron polarization was analyzed but the incident beam was unpolarized. We chose this setup partly for greater simplicity, but also because selecting the incident neutrons by Bragg reflection from a Si crystal gives a much smaller flux of higher harmonic reflections than the equivalent setup with a Heusler monochromator.

\section{Results}\label{s:LSCoO results}

\begin{figure}[!h]
\includegraphics*[scale=0.45,angle=0]{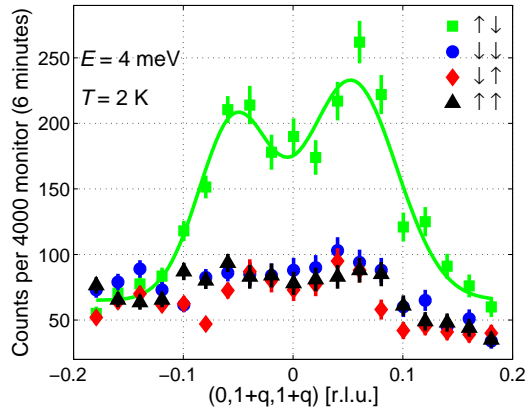}
\centering \caption{Polarized-neutron scattering from La$_{0.82}$Sr$_{0.18}$CoO$_{3}$. The Q-scans are measured at a fixed energy of
$E=4$\,meV with a fully-polarized setup. The sample temperature was $T=2$\,K, and a fixed final neutron energy of $E_{f}=14.7$\,meV was used. The four partial cross-sections correspond to the two possible spin states ($\uparrow$ or $\downarrow$) of the neutrons before and after scattering. The green line going through the points corresponding to the $\uparrow \downarrow$ signal is a guide to the eye.}\label{fig:LaSrCoO3 Fully polarized}
\end{figure}

Figure \ref{fig:LaSrCoO3 Fully polarized} shows data obtained with a fully-polarized setup. For neutron energy loss we measure magnon creation, $S_{-+}^{\rm M}$, and thus expect a signal for only one of the polarization configurations, as indeed is observed. Furthermore, we can also see that we were justified in using the half-polarized setup, since there is a clear difference between the sums $(S_{\uparrow \downarrow} + S_{\downarrow \downarrow})$ and $(S_{\uparrow \uparrow} + S_{\downarrow \uparrow})$.

\begin{figure}[!h]
\includegraphics*[scale=0.45,angle=0]{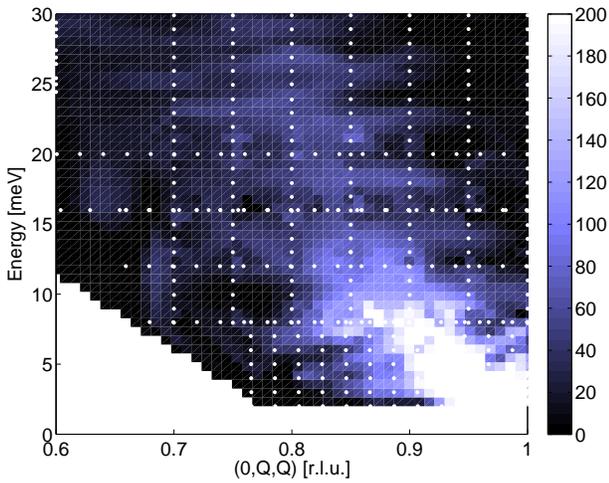}
\centering \caption[Colour map showing the magnetic
scattering]{Color map showing the intensity of the magnetic scattering ($S_{\circ\downarrow} - S_{\circ\uparrow}$) using the half-polarized
configuration. The figure shows a combination of all scans, with data from symmetrically equivalent wavevectors included. Data collected with different $E_{f}$ have been normalized to the same intensity scale. All measurements were taken with $T=2$\,K.}\label{fig:LaSrCoO3
Magnetic colourmap}
\end{figure}



The rest of the data were taken using the half-polarized setup. Figure \ref{fig:LaSrCoO3 Magnetic colourmap} shows all of the polarization-analyzed scans, i.e. $S_{\circ\downarrow} - S_{\circ\uparrow}$, combined. Scans taken with different $E_{f}$ have been normalized to the same scale by comparing several equivalent scans taken with different $E_{f}$ and interpolating the scale factors found. The colors, representing intensity, have been smoothed by interpolation, so the figure is for the purpose of illustration only. The white dots show points in ($\mathbf{Q},E$)--space where the scattering was measured, so special care should be taken when considering colors on this map far from any such dots, because the interpolation in such regions is a less reliable indicator of
the true intensity there. Note that some data have been
symmetrized, e.g. data taken at the wavevector $(0,1.4,1.4)$ are
shown on this figure at $(0,0.6,0.6)$ because they are
symmetrically equivalent once they have been corrected for the
magnetic form factor. This figure reveals a magnetic mode dispersing from the ferromagnetic wavevector $(0,1,1)$. As energy increases, the spin-waves broaden in both energy and wavevector quite significantly, and above about 20\,meV the magnetic scattering is very diffuse and hence very hard to characterize accurately.

The use of polarization analysis was essential to prove that the dispersing signal is indeed magnetic, and to map out the weak magnetic signal at higher energies. An optic phonon at 20\,meV, which appears in both $S_{\circ \uparrow}$ and $S_{\circ\downarrow}$, is successfully removed in the difference spectrum $S_{\circ\downarrow} - S_{\circ\uparrow})$.


\begin{figure}[!h]
\includegraphics*[scale=0.45,angle=0]{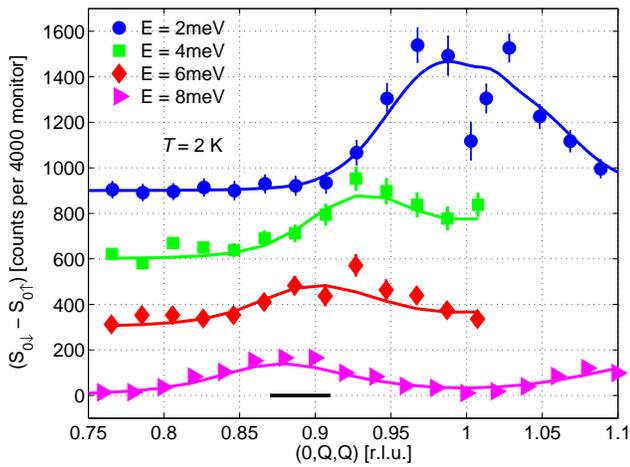}
\centering \caption[Constant energy Q-scans of the magnetic
scattering]{Constant energy Q-scans of the magnetic scattering
$(S_{\circ\downarrow} - S_{\circ\uparrow})$. Measurements were performed with $T=2$\,K and $E_{f}=14.7$\,meV. Successive scans are displaced by 300
for clarity. Fits to the data, described in
the main text, are shown as solid lines. The black horizontal line indicates the FWHM of the spectrometer's $\mathbf{Q}$
resolution, calculated using the RESTRAX software \cite{Restrax}.}\label{fig:LaSrCoO3 lowE Qscans Magnetic}
\end{figure}

Figure \ref{fig:LaSrCoO3 lowE Qscans Magnetic} shows several
fixed-energy Q-scans in the half-polarized ($S_{\circ\downarrow} - S_{\circ\uparrow}$) channel, which contains only magnetic scattering. Since all measurements were performed at $T=2$\,K the Bose thermal population factor $\{1 + n(E)\} \approx 1$ for the energy range considered here. Thus, the signal ($S_{\circ\downarrow} - S_{\circ\uparrow}$) is proportional to the generalized susceptibility $\chi_{\alpha \beta}''(\mathbf{Q},E)$ convoluted with the spectrometer's resolution. For the low energy Q-scans shown in figure \ref{fig:LaSrCoO3 lowE Qscans Magnetic} it was possible to use $E_{f}=14.7$\,meV, at which final energy the flipping ratio was significantly better and hence the cross-contamination of non-magnetic scattering into the magnetic scattering channel was much lower. A clear dispersion from the origin is visible, accompanied by a decrease in the intensity of the scattering.


\begin{figure}[!h]
\includegraphics*[scale=0.45,angle=0]{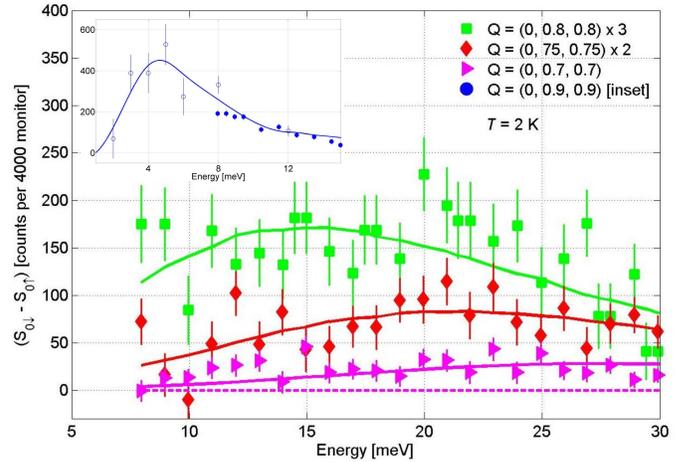}
\centering \caption[Fits to fixed-wavevector energy scans, using a
single Gaussian.]{Fixed-wavevector energy scans, from measurements with $E_{f}=34.8$\,meV. Measurements were performed with $T=2$\,K. Scans at $\mathbf{Q}=(0.8,0.8,0)$ and $\mathbf{Q}=(0.75,0.75,0)$ are multiplied by 3 and 2 respectively for clarity. Solid lines are
fits to the data, described in the main text, and the dashed line indicates the background level (fixed to zero). The inset shows a similar scan taken at $\mathbf{Q}=(0,0.9,0.9)$, with closed circles denoting data taken with the $E_{f}=34.8$\,meV configuration and open circles denoting data taken with $E_{f}=14.7$\,meV, rescaled by a factor of 2 for clarity. The solid line is a global fit to the data, described in the main text.}\label{fig:LaSrCoO3 Energy fits}
\end{figure}

Figure \ref{fig:LaSrCoO3 Energy fits} shows the magnetic signal ($S_{\circ\downarrow} - S_{\circ\uparrow}$) obtained from fixed-wavevector energy scans at a range of wavevectors away from the ferromagnetic zone center. Inset is shown an energy scan closer to the zone center, at $\mathbf{Q}=(0,0.9,0.9)$. The data show that as the wavevector increases from the ferromagnetic zone center, the
excitation disperses out to a higher energy. The data also show that, in energy, the excitations are extremely broad. There are two contributions towards the large width shown in these scans -- a steep dispersion and/or a large inverse lifetime for the excitations.

\section{Analysis and Discussion}\label{s:LSCoO discussion}

Our polarized neutron scattering data show clearly that the magnetic excitations in LSCoO are ferromagnetic in origin and are dispersive. The main question is whether the excitations arise from fluctuations of localized or itinerant magnetic moments. In both of these cases one would expect the dispersion to be approximated fairly well by a law quadratic in $q$ at low energies \cite{Squires spin waves,Lynn iron}. In metallic (i.e. itinerant electron) ferromagnets, however, the scattering only shows a peak in fixed-energy Q-scans, whereas it is quasi-elastic for fixed-wavevector energy-scans. In contrast the scattering from a Heisenberg
ferromagnet with localized magnetic moments has peaks in both energy-scans and Q-scans. It is clear from figure \ref{fig:LaSrCoO3 Energy fits}, especially when one takes note of the inset, that the scattering is not quasi-elastic. Rather the signal, though considerably broadened at wavevectors away from the ferromagnetic zone center, show peaks at non-zero energy. This suggests that the excitations are from fluctuations of localized magnetic moments, i.e. spin waves.

The data were fitted using a dispersion of the form,
\begin{equation}\label{eq:nn dispersion}
E_{q} = \Delta + 4SJ[3 - \rm{cos}(2\pi q_{h}) - \rm{cos}(2\pi q_{k}) - \rm{cos}(2\pi q_{l})],
\end{equation}

\noindent where $E_{q}$ is the energy of the excitation, $\mathbf{Q}=(q_{h},q_{k},q_{l})$ in reciprocal lattice units, $J$ is the nearest-neighbor exchange constant, $S=1/2$ is the spin of the Co$^{4+}$ ions, and $\Delta$ is an energy gap. Although it is not clear whether only the Co$^{4+}$ spins with $S=1/2$ are ferromagnetically correlated, or whether the Co$^{3+}$ spins with $S=1$ or $S=2$ are also correlated, it is possible
nevertheless to write an `average' Hamiltonian. In such a
Hamiltonian we assume an effective $S=1/2$ and then determine the exchange
constants accordingly. If there is no clear way to separate the
Co$^{4+}$ and the Co$^{3+}$ contributions to the magnetic
scattering then this is the only sensible approach to take. Note that the spin-wave spectrum extends between $\Delta$ and $\Delta + 16\mbox{J\,}S$ for dispersion from the $(0,1,1)$ wavevector parallel to $(0,1,1)$.

If the spin waves are damped then the measured intensity of the magnetic scattering $(S_{\circ\downarrow} - S_{\circ\uparrow})$ can
be described using a damped harmonic oscillator model, i.e.

\begin{equation}\label{eq: DHO equation}
(S_{\circ\downarrow} - S_{\circ\uparrow}) \propto \frac{4 \Gamma E E_{q}}{(E^{2} - E_{q}^{2})^{2} + 4
\Gamma^{2} E^{2}},
\end{equation}

\noindent where $\Gamma$ is the damping constant. We have data from both fixed-energy Q-scans and from fixed-Q energy-scans, and we were able to fit
these simultaneously to the same model (given by equation \ref{eq: DHO equation}) using the LIBISIS `multifit' software \cite{Libisis website}. In the fitting procedure equation \ref{eq: DHO equation} was convoluted with the spectrometer's resolution using the RESCAL software \cite{Rescal website}, and the background was fixed to zero. We first fitted scans where the neutron energy transfer was below 10\,meV, for which the dispersion can be approximated by a law quadratic in $q$, in order to determine a sensible upper limit for the exchange $J$. We then fitted all of the scans simultaneously with this constraint incorporated. Constraining the amplitude (i.e. the constant of proportionality in eq. \ref{eq: DHO equation}) to be the same for all scans produced the same fit parameters as when the amplitude was fitted independently for each scan. In the latter procedure, the variation in amplitude was found to be $\sim 20\%$, a factor which is broadly consistent with the size of variation expected due to variation in the vertical focussing of the Si monochromator as a function of neutron incident energy \cite{Nuc_inst_focus_paper}.

The lines on figures \ref{fig:LaSrCoO3 lowE Qscans Magnetic} and \ref{fig:LaSrCoO3 Energy fits} show the result of this fit. Notice that the fit for $E=2$\,meV shown in figure \ref{fig:LaSrCoO3 lowE Qscans Magnetic} does not give two resolved peaks, whereas visually the data appear to suggest that two peaks can be resolved. With the spectrometer resolution incorporated in the fitting, it was found not to be possible to get a fit with two resolved peaks without unrealistic settings for the spectrometer configuration.

\begin{figure}[!h]
\includegraphics*[scale=0.5,angle=0]{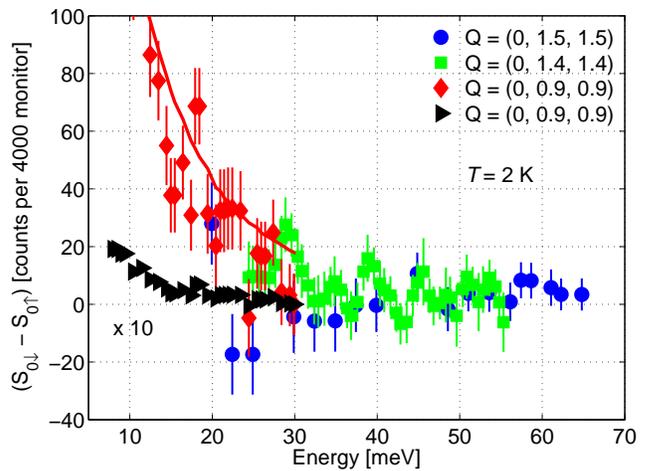}
\centering \caption[Comparison of low energy excitations with zone
boundary energy scans]{Fixed wavevector energy scans, comparing a
scan at $(0,0.9,0.9)$ where there is a clear spin-wave peak, with
scans at $(0,1.4,1.4)$ and $(0,1.5,1.5)$, where there is no
apparent signal. The scan at $(0,0.9,0.9)$ is shown twice, red diamonds and the red line show it to scale with a fit, whereas
black triangles show it reduced by a factor of
10. Data were taken at $T=2$\,K, and for all scans shown $E_{f}=34.8$\,meV was used.}\label{fig:LaSrCoO3 HighQ NoSignal}
\end{figure}

The parameters obtained from this fitting procedure were $J=6.5\pm0.2$\,meV and $\Delta=1.4\pm1$\,meV. The parameter $\Delta$ is required for the best fit, however we must point out that on a thermal neutron spectrometer such as IN20 it is difficult to prove conclusively the magnitude of a small gap such as this due to the resolution. Rather, one must use a cold-source neutron spectrometer to determine this accurately. Indeed, we find that within two standard deviations we measure the gap $\Delta$ to be zero. Furthermore, we found that in order to fit the data well, the damping parameter $\Gamma$ was required to increase with increasing energy. We set $\Gamma = \Gamma_{0} (E - \Delta)$, and the best fit was found with $\Gamma_{0}=0.7\pm0.1$. We found little change in the other fit parameters if we chose alternative forms for $\Gamma$, such as quadratic, cubic, etc., that give a steady increase. The large coefficient $\Gamma_{0}$ means that even for rather modest energies the excitations are very heavily damped, and would explain why there is no discernible signal in the fixed-wavevector energy-scans near the Brillouin zone boundary, shown in figure \ref{fig:LaSrCoO3 HighQ NoSignal}.

The results of the fitting can be compared to similar parameters in the CMR
manganite La$_{0.8}$Sr$_{0.2}$MnO$_{3}$ \cite{Moudden LSMO} (LSMO). For this material the spin-wave stiffness constant is $D_{\rm{LSMO}} \approx 150$\,meV\,\AA$^{2}$ and the energy gap is $\Delta_{\rm{LSMO}} \leq 0.04$\,meV. The spin-wave stiffness constant is obtained from the long-wavelength approximation of equation \ref{eq:nn dispersion}, so that $D = 2SJa^{2} = 94\pm 3$\,meV\,\AA$^{2}$ for LSCoO. It appears, then, that the low energy spin-waves are of comparable stiffness in LSCoO and LSMO, indicating that the ferromagnetic exchange between magnetic ions is also comparable.

In a Heisenberg model with only nearest-neighbor interactions the Curie temperature $T_{C}$ is related, through mean-field theory, to the exchange
energy by

\begin{equation}\label{eq:Tc mean field}
T_{C}^{\rm{calc}}=\frac{2z\mbox{J}_{1}S(S+1)}{3k_{\rm{B}}},
\end{equation}

\noindent where $z=6$ is the number of nearest neighbors, and all
the other symbols have their usual meanings. For $S=1/2$ and $\rm{J}=6.5$\,meV, this gives $T^{\rm{calc}}_{C}=227\pm5$\,K. This somewhat overestimates $T_{C}$ when compared with measurements of the susceptibility \cite{Kriener CaSrBa}, which indicate $T^{\rm{meas}}_{C}= 190$\,K. Equation \ref{eq:Tc mean field} does not take into account the effects of spin fluctuations. A fuller approach \cite{Rushbrooke Tc} which does include fluctuations gives

\begin{equation}\label{eq: modified Tc}
T_{C}^{\rm{calc}} = \frac{\mbox{J}_{1}}{k_{\rm{B}}}[2.90 S(S+1)-0.36],
\end{equation}

\noindent which would result in $T^{\rm{calc}}_{C}=137\pm3$\,K, which rather underestimates the measured value. However, for the Hamiltonian given in equation \ref{eq:nn dispersion} we have explicitly assumed that $S=1/2$, and in fact our measurements only allow us to determine the product $SJ$. If we assume that most of the magnetic response arises from the Co$^{3+}$ ions in a magnetic $S=1$ state, then equations \ref{eq:Tc mean field} and \ref{eq: modified Tc} would give Curie temperatures of 303\,K and 206\,K respectively.

Let us now consider the damping of the spin waves. First, we point out that given the relative steepness of the dispersion and the finite resolution of the spectrometer in $\mathbf{Q}-E$-space one might expect the measurements to be insensitive to any damping effects. In fact, we found that either setting the damping to be very small, or indeed keeping it constant, significantly reduced the quality of the fit compared to having damping that increases with energy. Interestingly, we also found that fits of very similar quality (as measured by the goodness-of-fit parameter $\chi^{2}$) could be achieved for $0.3<\Gamma_{0}<2$, with increases in the exchange $J$ accommodating increases in the damping gradient $\Gamma_{0}$. However, when only the scans taken at lower energies were considered this uncertainty in $\Gamma_{0}$ was removed and we converged on the best-fit values stated.

Broadening of spin waves has been observed in the manganites, and in different materials has been attributed to different mechanisms. These are (i) disorder in the mean magnetic structure \cite{Moussa everyone wrong}, so that there exists a distribution of values for J; (ii) there is a Stoner continuum of scattering above a certain energy, giving rise to very broad magnetic scattering; (iii) the magnons are scattered by phonons; or (iv) the magnons are scattered by electrons.

For the case of (i), the effect would have to be
significantly larger than that seen in La$_{1-x}$Ca$_{x}$MnO$_{3}$ to explain the level of damping observed here. We estimate that a distribution of J
with FWHM of $\sim5$\,meV would be required. A variation of this size would seem to be rather large for a bulk ferromagnet. However, the ferromagnetism does not become phase-pure until $x\geq 0.22$ \cite{He bulk meas prb}, and together with observations of incommensurate magnetic order \cite{Phelan ICM} and glassy behavior \cite{Wu glass phase} for $x<0.18$, there is some indication that variations in the exchange coupling are possible. If this were the case then one would expect that neutron scattering measurements of LSCoO with $x<0.18$ would reveal ferromagnetic excitations that were broader than those presented here, whereas for $x>0.18$ one would conversely expect the signal to be less strongly broadened.

In the cases of (ii) and (iii) one would expect to observe a sudden increase in the damping, corresponding to the energy of the lower boundary of the Stoner continuum, or the energy of the phonon respectively \cite{Zhang manganites review}. In the present case we observe a steady increase in $\Gamma$. Indeed if we repeat our fitting procedure for the fixed-energy Q-scans with independent values of $\Gamma$ for each scan we still observe a gradual increase with energy. This contrasts with measurements of, for example, La$_{0.7}$Ca$_{0.3}$MnO$_{3}$ \cite{Ye many manganites} where abrupt increases in a magnon's linewidth are observed when its dispersion crosses that of a phonon. On the other hand, the magnitude of the damping coefficient measured here is much larger than in the manganites, and the presence of a Stoner continuum would explain such heavy damping. In case (iv), a scenario which has been considered theoretically for various different manganites \cite{Golosov DE damping} using double exchange models, the damping parameter is expected to increase steadily, rather than suddenly. This bears some resemblance to our observations, however for many of the manganites for which this is considered in detail the magnitude of the damping is much smaller than that measured here.

\section{Conclusions}\label{s:LSCoO conclusions}

To conclude, our measurements suggest that the low-energy spin
excitations in LSCoO with $x=0.18$ can be described in terms of a simple localized Heisenberg ferromagnet. The spin wave stiffness was found
to be comparable to that found in similarly doped manganites. This spin wave stiffness was used to calculate the Curie temperature using a mean field model, assuming nearest-neighbor exchange and with and without corrections for spin fluctuations, and reasonable agreement was obtained with the measured value of $T_{\rm C}$. The inverse lifetime of the spin waves increases monotonically with increasing energy. We tentatively ascribe this effect as being due to either scattering of the magnons by electrons within a double-exchange framework, the presence of a Stoner continuum, or variations in the exchange parameter arising from microscopic phase separation.

We thank F. R. Wondre for x-ray Laue alignment of the sample, and T. G. Perring for helpful discussions. We are grateful for financial support from
the Engineering and Physical Sciences Research Council of Great Britain.

\section*{Appendix I: Pseudo-cubic notation}\label{a:rhombohedral_notation}

LSCoO crystallises in the rhombohedral space group $R\bar{3}c$ with lattice parameters $a=5.371$\AA\,\, and $\alpha=60.758^{\circ}$. This lattice can
be regarded as a slightly distorted cubic lattice, with pseudo-cubic lattice parameters $a\approx 3.8\AA$ and $\alpha = 90^{\circ}$. The conversion
from rhombohedral to cubic co-ordinates in real space is given by

\begin{eqnarray}
\left[1 0 0 \right]_{\rm{R}} &=& \left[1 0 1 \right]_{\rm{C}}\nonumber \\
\left[0 1 0 \right]_{\rm{R}} &=& \left[1 1 0 \right]_{\rm{C}}\\
\left[0 0 1 \right]_{\rm{R}} &=& \left[0 1 1 \right]_{\rm{C}}\nonumber
\label{eq:rhombohedral real space}
\end{eqnarray}

\noindent where the subscripts R and C refer to rhombohedral and cubic lattices respectively. In reciprocal space, because the rhombohedral axes are
not orthogonal, this leads to

\begin{eqnarray}
(1, 0, 0)^{\ast}_{\rm{R}} &=& (\frac{1}{2}, \overline{\frac{1}{2}}, \frac{1}{2})^{\ast}_{\rm{C}}\nonumber \\
(0, 1, 0)^{\ast}_{\rm{R}} &=& (\frac{1}{2}, \frac{1}{2}, \overline{\frac{1}{2}})^{\ast}_{\rm{C}}\\
(0, 0, 1)^{\ast}_{\rm{R}} &=& (\overline{\frac{1}{2}}, \frac{1}{2}, \frac{1}{2})^{\ast}_{\rm{C}}\nonumber
\label{eq:rhombohedral real space}
\end{eqnarray}
\newline


\end{document}